\documentclass[12pt]{article}

\ifx\pdfoutput\undefined
\usepackage[dvips,bookmarks]{hyperref}  
\else
\usepackage{hyperref}   
\fi

\usepackage{graphicx}

\hypersetup{colorlinks,bookmarksopen,bookmarksnumbered,citecolor=blue,
        pdfstartview=FitH}
\def\hhref#1{\href{http://arxiv.org/abs/hep-th/#1}{hep-th/#1}} 
\def\mhref#1{\href{mailto:#1}{#1}}              

\def\bop#1{\setbox0=\hbox{$#1M$}\mkern1.5mu
    \vbox{\hrule height0pt depth.04\ht0
    \hbox{\vrule width.04\ht0 height.9\ht0 \kern.9\ht0
    \vrule width.04\ht0}\hrule height.04\ht0}\mkern1.5mu}
\def\bo{{\mathpalette\bop{}}}                        
\def\frac#1#2{{\textstyle{#1\over#2}}}     

\def\on#1#2{{\buildrel{\mkern2.5mu#1\mkern-2.5mu}\over{#2}}}
\def\fzero{\on\circ F{}}

\parskip=\medskipamount         
\thicklines             
\begin{document}

\begin{titlepage}
\setcounter{page}{0} 
\begin{flushright}
YITP-SB-04-48\\ September 15, 2004\\
\end{flushright}

\begin{center}
{\centering {\LARGE\bf Gauge-covariant S-matrices\\
 for field theory and strings \par} }

\vskip 1cm {\bf Haidong Feng\footnote{E-mail address:
\mhref{hfeng@ic.sunysb.edu}}, Warren Siegel\footnote{E-mail
address: \mhref{siegel@insti.physics.sunysb.edu}}} \\ \vskip 0.5cm

{\it C.N. Yang Institute for Theoretical Physics,\\
 State University of New York, Stony Brook, 11790-3840 \\}

\end{center}

\begin{abstract}
S-matrices can be written Lorentz covariantly in terms of free field strengths for vector states, allowing arbitrary gauge choices.  In string theory the vertex operators can be chosen so this gauge invariance is automatic.  As examples we give four-vector (super)string tree amplitudes in this form, and find the field theory actions that give the first three orders in the slope.
\end{abstract}

\end{titlepage}

\section{Introduction}

\subsection{Gauges}

An interesting feature of four-point amplitudes with four external gauge fields in both $D=10$ superstrings and maximally supersymmetric gauge theories in $D\le 10$ (and by supersymmetry, arbitrary external massless states) is that the kinematic factors are identical at the tree and one-loop level \cite{gs}.  Because lower-point amplitudes vanish in these theories, the one-loop four-point amplitude consists of one-particle-irreducible graphs in the field theory case, and is thus expressed directly in terms of field strengths as a contribution to the effective action in a background-field gauge calculation, as the ``non-field-strength" contributions (from non-spin couplings) exactly cancel \cite{REALgs}.  On the other hand, tree graphs are never expressed in terms of field strengths, so the identity of these kinematic factors seems somewhat mysterious.

In general field theories, the fact that S-matrices always have external propagators amputated means that the generating functional for the S-matrix (as opposed to that for Green functions) can always be expressed in terms of fields rather than sources \cite{F}.  (Consider, e.g., the external vector states for the tree amplitude of an electron in an external electromagnetic field.)  In fact, the external line factors of Feynman graphs are (asymptotic) fields, and satisfy their (free) wave equations and (linear) gauge conditions.  However, the gauge conditions imposed on external states generally do not match those applied to internal ones, neither for propagators in loops (``quantum gauge") nor in attached trees (``background gauge"):  Usually the latter two gauges are some variation of the Fermi-Feynman gauge, while the external states satisfy a Landau gauge, further restricted to some type of unitary gauge (lightcone or Coulomb) by the residual gauge invariance.  An exception is when external polarizations are summed over in a cross section, a procedure that is often more cumbersome because cross sections involve double sums (i.e., over both amplitudes and their complex conjugates).

The consistency of this procedure follows from the fact that in general {\it three} independent gauges can be chosen in the calculation of an S-matrix element from Feynman diagrams, corresponding to three steps in the procedure:  (1) First calculate the effective action, using the background field method.  The gauge for the ``quantum" fields, which appear inside the loops, is fixed but the background fields are not gauge-fixed.  The resulting effective action, which depends only on the background fields, is thus gauge invariant, not merely BRST invariant (and in fact is not a functional of the ghosts).  (2) Calculate the generating functional for the S-matrix from ``tree" graphs of the effective action, treating the full effective action as ``classical", fixing the gauge for the (background) fields of the effective action.  The result can always be expressed as a functional of linearized, on-shell field strengths only, in a Lorentz and gauge covariant way.  (3) Calculate a specific S-matrix element, choosing a (linear) unitary gauge condition for the external gauge fields, or expressing the external field strengths directly in terms of polarizations.  

It is the second step that will be the focus of this paper.  We will also examine its analog in string and superstring theory.  In that case, with the usual first-quantized methods, the effective action does not appear, so the procedure reduces to two steps:  (1) Calculate the S-matrix in terms of field strengths by using {\it gauge-covariant vertex operators} \cite{fs}.  (2) Same as step 3 of the field theory case.  The main difference in the string case is that gauge invariance at the next-to-last step is automatic (although there is still some work to rearrange the result in terms of field strengths).  The advantages of having the third gauge invariance are similar to those of the other gauge invariances, since the result (a) can be applied to different gauges (e.g., lightcone or Coulomb), depending on the application, (b) is generally simpler, since various terms of various derivatives of gauge fields can be combined into field strengths, (c) is more unique, simplifying comparison of different contributions, and (d) is manifestly Lorentz covariant.  

Some of these advantages can also be obtained by instead using a twistor formalism (``spinor helicity" \cite{sh}, ``spacecone" \cite{cs}, etc.), but that approach does not generalize conveniently to higher dimensions.  In fact, the two methods are somewhat related in $D=4$.  As an example, consider the ``maximally helicity violating" tree amplitudes of Yang-Mills theory \cite{pt}:  In the usual twistor notation, these are written as
$$ {\cal A} =  {\langle ij \rangle^4 \over \langle 12 \rangle \langle 23 \rangle \dots
\langle n1 \rangle} , \quad\quad  
\langle kl \rangle  = \lambda_k^\alpha \lambda_{l\alpha} $$
for an $n$-point amplitude with $i$ and $j$ labeling the lines whose helicites differ from the rest.  The twistors themselves are ``square roots" of the momenta,
$$ p_\alpha{}^{\dot\alpha} = \lambda_\alpha \bar\lambda^{\dot\alpha} $$
so no residue of gauge invariance is visible, but manifestation of Lorentz invariance is possible because in $D=4$ the little group is just U(1), as represented by helicity.  On the other hand, a twistor can also be interpreted as the square root of (the selfdual $f$ or anti-selfdual $\bar f$ part of) an antisymmetric tensor:  In an appropriate normalization for external lines,
$$ f_\alpha{}^\beta = \lambda_\alpha \lambda^\beta , \quad\quad
\bar f_{\dot\alpha}{}^{\dot\beta} = \bar\lambda_{\dot\alpha} \bar\lambda^{\dot\beta} $$
as follows from Maxwell's equations.  Thus the result can easily be expressed in terms of field strengths and the usual (helicity-independent) momentum invariants by completing the denominator of the amplitude to the square of its absolute value (thus making the usual pole structure obvious):  In 2$\times$2 matrix notation,
$$ {\cal A} = {tr(f_i f_j)tr(p_j \bar f_{j+1}...\bar f_n\bar f_1...\bar f_{i-1}p_i^*)
tr(p_i\bar f_{i+1}...\bar f_{j-1}p_j^*) \over p_1\cdot p_2...p_n\cdot p_1} $$

In string theory, the gauge-boson vertex operator $A(X) \cdot\partial X$, expanded in plane waves as $A(X) = \epsilon e^{ik \cdot X}$, is not gauge covariant, and requires the gauge condition $\partial\cdot A=0$ for worldsheet conformal invariance.  In a previous paper \cite{fs} we derived gauge-covariant vertex operators for (super)strings and used them to calculate the three-vector vertex:  The result was the cubic term from the gauge-unfixed $F^2$ Yang-Mills action (and in the bosonic string, also an $F^3$ term).  In this paper, we will use this gauge-covariant vertex operator to compute the gauge-invariant tree amplitude
between 4 gauge bosons.  In particular, to our knowledge a complete, explicit  expression for this amplitude (i.e., not simply as a functional derivative of some generating functional) in bosonic string theory has not appeared previously in the literature.  Then we will reproduce the same amplitudes at order 1, $\alpha'$, and $\alpha'^2$ from the appropriate $F^2$, $F^3$ (for the bosonic string), and $F^4$ terms in a field theory action.

\subsection{Results}

For the bosonic string we find the amplitude (see subsection \ref{subsection1}):
\begin{equation}\label{boson amp}
( K_0 + \alpha' K_1 + \alpha'^2 stu K_2)\ \alpha'^2\ 
{\Gamma(-\alpha' s) \Gamma(-\alpha' t) \over \Gamma(1-\alpha' s-\alpha' t)}
\end{equation}
where we have factored out the usual coupling constants and momentum conservation $\delta$-function, as well as Chan-Paton factors for cyclic ordering.  The kinematic factors are
\begin{eqnarray}\label{K_0}
K_0 & = & (4\fzero^1_{\mu \nu} \fzero^2_{\nu \sigma}
\fzero^3_{\sigma \rho} \fzero^4_{\rho \mu} -
\fzero^1_{\mu \nu} \fzero^2_{\nu \mu}
\fzero^3_{\sigma \rho} \fzero^4_{\rho \sigma}) + 2 ~ permutations \nonumber \\
& \equiv & t^{\mu \nu \rho \sigma \alpha \beta \gamma \delta}
\fzero^1_{\mu \nu} \fzero^2_{\rho \sigma}
\fzero^3_{\alpha \beta} \fzero^4_{\gamma \delta} ,
\end{eqnarray}
\begin{equation}
K_1 = [ 4(\fzero^1_{\mu \nu} \fzero^4_{\nu \mu})
(k^1 - k^4)_{\tau} \fzero^2_{\tau \sigma}
\fzero^3_{\sigma \lambda} k^4_{\lambda} +
8\fzero^1_{\nu [ \mu} \fzero^2_{\sigma ] \nu }
\fzero^3_{\sigma\rho} k^4_\rho \fzero^4_{\mu\tau}
k^1_\tau ]  + 3 ~ permutations
\end{equation}
\begin{equation}
K_2 = -2 \left(
{\fzero^1_{\mu \nu} \fzero^4_{\nu \mu}
\fzero^2_{\sigma \rho} \fzero^3_{\rho \sigma} \over t(1+\alpha' t)}  
+ {\fzero^1_{\mu \nu} \fzero^2_{\nu \mu}
\fzero^3_{\sigma \rho} \fzero^4_{\rho \sigma} \over s(1 +\alpha' s)} 
+ {\fzero^1_{\mu \nu} \fzero^3_{\nu
\mu} \fzero^2_{\sigma \rho} \fzero^4_{\rho
\sigma} \over u(1+\alpha' u)} \right)
\end{equation}
Here, the permutations in $K_0$ are the order $1342$ and $1423$
which replace the cyclic order $1234$ and the 3 permutations in
$K_1$ are the replacing of $1234$ by $2341$, $3412$ and $4123$. We
also have the definitions 
$$\fzero^i_{\mu \nu} =
k^i_{[ \mu} \epsilon^i_{\nu ]} =
k^i_{\mu} \epsilon^i_{\nu} - k^i_{\nu} \epsilon^i_{\mu}$$
 and
\begin{equation}
s = - (k^1 + k^2)^2 , ~~~ t = - (k^1 + k^4)^2 , ~~~ u = - (k^1 +k^3)^2.
\end{equation}
Because $\bo\fzero=0$ is gauge invariant, we can always set any $(k^i)^2=0$ once all external line factors have been written in terms of $\fzero$'s. 
The $K_2$ term in (\ref{boson amp}) can be regarded as the
contribution of tachyon poles in the $s$ and $t$ channels (the apparent $u$ pole is canceled by the $\Gamma$'s), and will be absent in the 
corresponding superstring amplitude, while the $K_1$ term corresponds to the contribution from an $F^3$ term in the field theory action, and hence is 
also absent in the presence of supersymmetry. 
(These amplitudes agree with earlier results obtained in the Landau gauge \cite{klt}.)

Expanding this amplitude in orders of $1$, $\alpha'$ and
$\alpha'^2$, it follows from the classical gauge theory action (see subsection \ref{subsection2}):
\begin{eqnarray}\label{bosonaction}
S  = \frac{1}{g_{YM}^{2}} \int d^D x\  [ \hskip-1.5em && - 
\frac{1}{4}
Tr(F^{\mu \nu} F_{\mu \nu})  - \frac{2i\alpha'}{3}
Tr({F_{\mu}}^{\nu} {F_{\nu}}^{\omega} {F_{\omega}}^{\mu} )
\nonumber \\ && - \frac{\pi^2 \alpha'^2}{4!} t^{\mu \nu \rho
\sigma \alpha \beta \gamma \delta} Tr(F_{\mu \nu} F_{\rho \sigma}
F_{\alpha \beta} F_{\gamma \delta})  \nonumber \\ &&
+\frac{\alpha'^2}{2} Tr( F_{\mu \nu} F_{\nu \mu} F_{\rho \sigma}
F_{\sigma \rho} - F_{\mu \nu} F_{\rho \sigma} F_{\nu \mu}
F_{\sigma \rho} ) ].
\end{eqnarray}

In the Neveu-Schwarz sector of the superstring, the 4-point tree amplitude is
\begin{equation}
K_0\ \alpha'^2\ {\Gamma(-\alpha' s) \Gamma(-\alpha'
t) \over \Gamma (1-\alpha' s - \alpha' t)} 
\end{equation}
with the same $K_0$ as defined in the bosonic case. Then, the low
energy limit in $O(\alpha'^0)$, $O(\alpha')$ and $O(\alpha'^2)$
follows from the classical action (see section \ref{section3}):
\begin{equation}\label{superaction}
S  = \frac{1}{g_{YM}^{2}} \int d^D x ~ [  - \frac{1}{4}
Tr(F^{\mu \nu} F_{\mu \nu})  - \frac{\pi^2 \alpha'^2}{4!} t^{\mu
\nu \rho \sigma \alpha \beta \gamma \delta} Tr(F_{\mu \nu} F_{\rho
\sigma} F_{\alpha \beta} F_{\gamma \delta}) ] .
\end{equation}
(These actions agree with those obtained from non-gauge-covariant amplitudes \cite{others}.)

\section{Bosonic string}

\subsection{Four-point amplitudes}\label{subsection1}

For the bosonic string, as in the previous paper, using the BRST
operator
\begin{equation}
Q = \oint \frac{1}{2\pi i} dz (-\frac{1}{4 \alpha'} c
\partial X \cdot \partial X + b c \partial c )
\end{equation}
and the integrated vertex operator for gauge bosons
\begin{equation}\label{inter vertex}
\oint W = \oint A \cdot \partial X
= \oint \epsilon \cdot \partial X e^{ik \cdot X} ,
\end{equation}
we found an unintegrated BRST invariant vertex operator without
gauge fixing
\begin{equation}\label{gauge vertex}
V = c A \cdot \partial X -\alpha'
(\partial c) \partial \cdot A 
= c \epsilon \cdot \partial X e^{ik \cdot X} - \alpha'
(\partial c) ik \cdot\epsilon e^{ik \cdot X} .
\end{equation}

Using the integrated vertex operator $\oint W$ (\ref{inter
vertex}) and the gauge invariant vertex operator $V$ defined in
(\ref{gauge vertex}), the gauge invariant N-point amplitude for
gauge bosons can be constructed in the bosonic string.
Specifically, The 4-point amplitude is:
\begin{eqnarray}\label{4-point amp}
{\cal A}_4 & = & \frac{g_{YM}^2}{2\alpha'^2} \langle V(y_1) \int d
y_2 W(y_2) V(y_3) V(y_4) \rangle
\end{eqnarray}

In the upper-half complex plane, the $X$ propagator is $-2\alpha' ln
|z'-z| \eta^{\mu \nu}$ and
\begin{eqnarray}
& \langle c(y_1) c(y_2) c(y_3) \rangle = y_{12} y_{13} y_{23}
\nonumber \\ & \langle \partial_{y_1} c(y_1) c(y_2) c(y_3) \rangle =
\partial_{y_1} ( y_{12} y_{13} y_{23}), \cdots .
\end{eqnarray}
Conventionally, set $y_1 = 0, y_3=1, y_4 \rightarrow \infty$ and
integrate $y_2$ from $0$ to $1$.

To write the 4-point amplitude in a gauge covariant form, the {\it
gauge-invariant} equation of motion of the free vector is necessary:
\begin{equation}\label{motion equ}
\partial^\mu F_{\mu\nu} = 0 \quad \quad
or \quad \quad k^2 \epsilon^{\mu} - k^{\mu} (k \cdot \epsilon) = 0
\end{equation}

Notice
\begin{equation}
  \int_0^1 dy ~ y^{a} (1-y)^{b} = \frac{\Gamma(a+1)
  \Gamma(b+1)}{\Gamma(a+b+2)}
\end{equation}
with
\begin{equation}
  \Gamma(a) = \int^{\infty}_{0} dt ~ t^{a-1} e^{-t}, ~~~~~~~
  \Gamma(a+1) = a \Gamma(a).
\end{equation}

For $y_4 \rightarrow \infty$ and $k^1 +k^2 +k^3 +k^4 =0$, the
factor appearing in ${\cal A}_4$
\begin{equation}
|y_{14}|^{2\alpha' k^1 \cdot k^4} |y_{24}|^{2\alpha' k^2 \cdot
k^4} |y_{34}|^{2\alpha' k^3 \cdot k^4} \rightarrow
|y_4|^{-2\alpha' k^4 \cdot k^4}.
\end{equation}
Using the equation of motion (\ref{motion equ}), this factor is
just $1$ if the amplitude is written in a gauge-covariant form.
Finally, the amplitude between 4 gauge bosons is given by eq.\ (\ref{boson amp}).

Taking the expansion in $\alpha'$
\begin{equation}\label{gamma expansion}
{\Gamma(-\alpha' s)
\Gamma(-\alpha' t) \over \Gamma(1-\alpha' s-\alpha' t)} =
{1 \over \alpha'^2 st} - {\pi^2 \over 6} + O(\alpha'),
\end{equation}
the lower orders till $O(\alpha'^2)$ of the amplitude ${\cal A}_4$ are
\begin{equation}\label{low order}
{K_0 + \alpha' K_1 \over st} + \alpha'^2
(-\frac{\pi^2}{6} K_0 + uK'_2) ,
\end{equation}
with
\begin{equation}
K'_2 = -2 \left( 
{ \fzero^1_{\mu \nu}
\fzero^2_{\nu \mu} \fzero^3_{\sigma \rho}
\fzero^4_{\rho \sigma} \over s} +
{ \fzero^1_{\mu \nu} \fzero^4_{\nu \mu}
\fzero^2_{\sigma \rho} \fzero^3_{\rho \sigma} \over t} +
{ \fzero^1_{\mu \nu}
\fzero^3_{\nu \mu} \fzero^2_{\sigma \rho}
\fzero^4_{\rho \sigma} \over u} \right)
\end{equation}

\subsection{Action in effective Yang-Mills theory}\label{subsection2}

\begin{figure}
\begin{center}
\includegraphics[width=10.5cm,height=4.5cm]{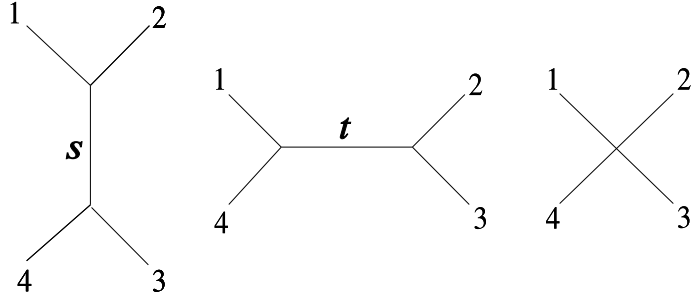}
\end{center}
\caption{\label{fig1} The $s$- and $t$-channel diagrams for 4 gauge
bosons coupled by $F^2$ vertices in $S_1$.}
\end{figure}

Clearly, the amplitude in $O(\alpha'^{0})$ in (\ref{low
order}) corresponds to the 4-point amplitude from 3 Feynman
diagrams in Yang-Mills theory
\begin{equation}\label{YM interaction}
S_1 = \frac{1}{g_{YM}^{2}} \int d^D x [ -\frac{1}{4} Tr(F^{\mu
\nu} F_{\mu \nu}) ] ,
\end{equation}
as shown in Fig.~\ref{fig1}.

\begin{figure}[tbh]
\begin{center}
\includegraphics[width=10.5cm,height=7.5cm]{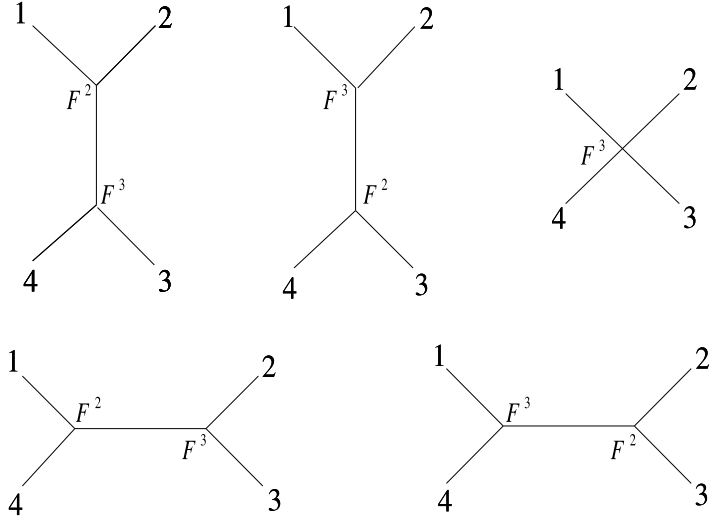}
\end{center}
\caption{\label{fig2} The $s$- and $t$-channel diagrams for 4 gauge
bosons coupled by one $F^2$ vertex in $S_1$ and one $F^3$ vertex
in $S_2$.}
\end{figure}

For the bosonic case, as mentioned in the previous paper, there is
a cubic interaction for gauge bosons:
\begin{equation}\label{cubic term}
S_2 = \frac{-2i\alpha'}{3g_{YM}^{2}} Tr({F_{\mu}}^{\nu}
{F_{\nu}}^{\omega} {F_{\omega}}^{\mu} )
\end{equation}
Thus, in the field theory to $O(\alpha')$ there are 5
Feynman diagrams for the 4-point amplitude, as shown in
Fig.~\ref{fig2}.

The summation of amplitudes from these 5 diagrams is
\begin{eqnarray}\label{amp of field}
-2 \alpha' ( K'_1 + 3 ~ permutations )
\end{eqnarray}
in which $K'_1$ is
\begin{equation}
\frac{p_a}{p^2} (\fzero^4_{ab} \fzero^1_{b\tau} -
\fzero^1_{ab} \fzero^4_{b\tau}) [2 \epsilon^2_{\tau}
(\epsilon^3 \cdot k^2) - 2 \epsilon^3_{\tau} (\epsilon^2 \cdot
k^3) + (k^3 - k^2)_{\tau} ( \epsilon^2 \cdot \epsilon^3) ] +
\fzero^1_{ab} \fzero^2_{bc} (\epsilon^3_c
\epsilon^4_a - \epsilon^3_a \epsilon^4_c)
\end{equation}
and $p= - k^1 - k^4$. The 3 permutations are the replacing of
$1234$ by $2341$, $3412$ and $4123$ in $K'_1$. To rewrite it in
gauge covariant form, apply the gauge transformation
\begin{equation}
  \epsilon^i_{\mu} \rightarrow \epsilon^i_{\mu} - k^i_{\mu}
  {\epsilon^i \cdot k^{i+1} \over k^{i} \cdot k^{i+1}} =
  - {k^{i+1}_{\nu} \fzero^i_{\mu \nu} \over k^i \cdot k^{i+1}} ,
\end{equation}
where $i+1\to1$ for $i =4$. Then, by using the {\it
gauge-invariant} equation of motion of the free vector
(\ref{motion equ}) and the Bianchi identity
\begin{equation}
k_{[ \mu} \fzero_{\nu \sigma ]} = 0
\end{equation}
the amplitude (\ref{amp of field}) is exactly the same as 
$O(\alpha')$ in the amplitude ${\cal A}_4$ in (\ref{low order}). It
agrees with the existence of $F^3$ terms in the Lagrangian density as
predicted by the three-point amplitude in the previous paper.

It is known that the superstring predicts a
higher-derivative gauge interaction $F^4$. In the bosonic case,
$O(\alpha'^2)$ in (\ref{boson amp}) will give
similar interactions.

\begin{figure}
\begin{center}
\includegraphics[width=10.5cm,height=4.5cm]{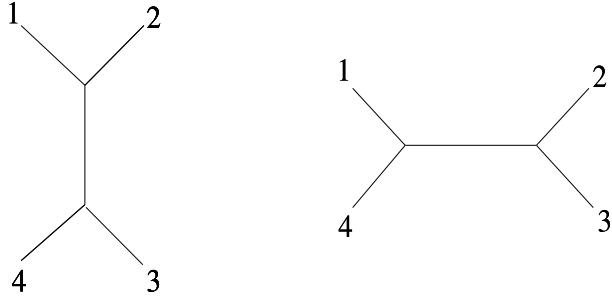}
\end{center}
\caption{\label{fig3}The $s$- and $t$-channel diagrams for 4 gauge
bosons coupled only by the $F^3$ vertex. }
\end{figure}

Because there is a cubic interaction $S_2$ in (\ref{cubic term}),
two Feynman diagrams, as shown in Fig.~\ref{fig3}, will give
directly a gauge covariant amplitude in $O(\alpha'^2)$
\begin{equation}
\alpha'^2 \left( {s-u\over t} \fzero^1_{\mu
\nu} \fzero^4_{\nu \mu} \fzero^2_{\sigma \rho}
\fzero^3_{\rho \sigma} + {t-u\over s}
\fzero^1_{\mu \nu} \fzero^2_{\nu
\mu} \fzero^3_{\sigma \rho} \fzero^4_{\rho \sigma} \right)
\end{equation}

But this is not equal to the $O(\alpha'^2)$ part of the string amplitude
(\ref{low order}). The difference between them represents
higher-derivative interactions, i.e., the $F^4$ interactions in
the effective theory. The difference is composed of two parts. One
is
\begin{equation}
B_1 = -\frac{\pi^2}{6} \alpha'^2 K_0,
\end{equation}
and the other is
\begin{equation}
B_2 = \alpha'^2 (\fzero^1_{\mu \nu} \fzero^4_{\nu \mu}
\fzero^2_{\sigma \rho} \fzero^3_{\rho \sigma} +
\fzero^1_{\mu \nu} \fzero^2_{\nu \mu}
\fzero^3_{\sigma \rho} \fzero^4_{\rho \sigma} - 2
\fzero^1_{\mu \nu} \fzero^3_{\nu \mu}
\fzero^2_{\sigma \rho} \fzero^4_{\rho \sigma})  .
\end{equation}

To convert the amplitude to the Lagrangian density, replace
$\fzero_{\mu \nu}$ by $- i F_{\mu \nu}$ and include
a factor of $1/4$ for the cyclic identity (as well as the usual overall factor $1/g_{YM}^2$). So from $B_1$ we
obtain
\begin{equation}\label{power4 interaction}
-\frac{\pi^2 \alpha'^2}{4! g_{YM}^2} t^{\mu \nu \rho \sigma \alpha
\beta \gamma \delta} Tr(F_{\mu \nu} F_{\rho \sigma} F_{\alpha
\beta} F_{\gamma \delta}) ,
\end{equation}
which we will see is same as the one from the
superstring, while using the same method, from $B_2$ we obtain
\begin{equation}
\frac{\alpha'^2}{2g_{YM}^2} Tr( F_{\mu \nu} F_{\nu \mu} F_{\rho
\sigma} F_{\sigma \rho} - F_{\mu \nu} F_{\rho \sigma} F_{\nu \mu}
F_{\sigma \rho} ) ,
\end{equation}
which is absent in the superstring case.
The low energy limit (\ref{low order}) of amplitude
${\cal A}_4$ in (\ref{boson amp}) then corresponds to the effective
action given in eq.\ (\ref{bosonaction}).

\section{Superstring}\label{section3}

In the case of the Neveu-Schwarz sector of the Ramond-Neveu-Schwarz formulation of the  superstring, the language of the ``Big
Picture"~\cite{bigpicture} will be used. Define 
$$ Z = (z, \theta) , \quad
X^{\mu}(Z) = x^{\mu} (z) + i \theta \psi (z) , \quad 
C = c + \theta \gamma , \quad
D_{\theta} = \partial_{\theta} + \theta \partial_z $$
where $c$ and $\gamma$ are the anticommuting and commuting superconformal
ghosts.  As mentioned in
our previous paper, the integrated vertex operator is 
$$ \oint W =
\oint A(X) \cdot D_{\theta} X =
\oint \epsilon \cdot D_{\theta} X e^{ik \cdot X} $$
Then the
BRST invariant vertex operator is found in the commutator as 
$$ [Q, W\} = D_{\theta} V $$
where $Q$ is the BRST operator. To simplify the calculation, we choose units $\alpha'=2$; $\alpha'$ will be restored in the final result by the replacements
$\epsilon\to\sqrt{\alpha'/2}\thinspace \epsilon$
and $k\to\sqrt{\alpha'/2}\thinspace  k$. Then,
\begin{eqnarray}\label{nsr vertex}
V & = & - D_{\theta} [C (\epsilon \cdot D_{\theta} X) e^{ik \cdot X
(Z) } ] + \frac{1}{2} (D_{\theta} C) (D_{\theta} X \cdot \epsilon)
e^{ik \cdot X (Z) } \nonumber \\ & & - 2 i (\epsilon \cdot k)
(\partial C) e^{ik \cdot X (Z)} .
\end{eqnarray}
In this convention, the propagator
\begin{equation}\label{propagator}
 X^{\mu}(z',
\theta') X^{\nu}(z, \theta) \sim -4 ln |z'- z - \theta' \theta |
\eta^{\mu \nu}
\end{equation}
and the correlation function
\begin{eqnarray}
& & \langle 0 | C(z_1, \theta_1) C(z_2, \theta_2) C(z_3, \theta_3) |
0 \rangle \nonumber \\ & = & \theta_1 \theta_2 z_3 (z_1 + z_2) +
\theta_2 \theta_3 z_1 (z_2 + z_3) + \theta_3 \theta_1 z_2 (z_3 +
z_1)
\end{eqnarray}

Then the 4-point amplitude in the  superstring can be written as
\begin{eqnarray}\label{4-point amp in NSR}
{\cal A}_4^{NSR} & = & -\frac{2g_{YM}^2}{\alpha'^2} \langle V(Z_1)
\int d z_2 d\theta_2 W(Z_2) V(Z_3) V(Z_4) \rangle,
\end{eqnarray}
with $z_1 = 0, z_3= 1, z_4 \rightarrow \infty$ and integrating
$z_2$ from $0$ to 
$1$.

The vertex operator (\ref{nsr vertex}) can also be written as
\begin{eqnarray}\label{new nsr vertex}
V & = & - \frac{1}{2} (D_{\theta} C) (\epsilon \cdot D_{\theta} X)
e^{ik \cdot X (Z) } + C D_{\theta} [( D_{\theta} X \cdot \epsilon)
e^{ik \cdot X (Z)}] \nonumber \\ & & - 2 i (\epsilon \cdot k)
(\partial C) e^{ik \cdot X (Z)} .
\end{eqnarray}

Using the anticommutation relation between $C$, $D_{\theta}$,
and $\int d\theta$, move $C$, $D_{\theta} C$, and $\partial C$ to the
left side of ${\cal A}_4^{NSR}$. To make the calculation simpler,
we first set $\theta_1$, $\theta_3$, and $\theta_4$ to zero. Thus we only have to
compute the terms independent of $\theta_1$, $\theta_3$, and $\theta_4$.
So the amplitude comes only from the parts with two $D_{\theta}
C$'s and one $C$ or $\partial C$.

For the same reason as in the bosonic case, the factor
$$ |y_{14}|^{2\alpha' k^1 \cdot k^4} |y_{24}|^{2\alpha' k^2 \cdot
k^4} |y_{34}|^{2\alpha' k^3 \cdot k^4} $$ 
appearing in ${\cal A}_4^{NSR}$ is just $1$ if the rest of the amplitude can be written in gauge-invariant form.

After restoring $\sqrt{\alpha'/2}$'s, we find
\begin{equation}\label{amp of NSR}
{\cal
A}_4^{NSR} (\theta_1=0,\theta_3=0,\theta_4=0) =
\alpha'^2 K_0 {\Gamma(-\alpha' s) \Gamma(-\alpha'
t) \over \Gamma (1-\alpha' s - \alpha' t)}
\end{equation}
where $K_0$ is defined by (\ref{K_0}).

To check independence from our choice $\theta_1=\theta_3=\theta_4=0$, we look at conformal invariance of the amplitude.
Since the vertex operator $V$ has the weight $\alpha' k^2$, the
4-point amplitude transforms as
\begin{eqnarray}
& & \langle V'(z'_1,\theta'_1) \oint W V'(z'_3,\theta'_3)
V'(z'_4,\theta'_4) \rangle \nonumber \\ & = & (D_{\theta_1}
\theta'_1)^{-2\alpha' k_1^2} (D_{\theta_3} \theta'_3)^{-2\alpha'
k_3^2} (D_{\theta_4} \theta'_4)^{-2\alpha' k_4^2} \langle
V(z_1,\theta_1) \oint W V(z_3,\theta_3) V(z_4,\theta_4) \rangle\quad\quad
\end{eqnarray}
Through a conformal transformation $\theta_1=0 \rightarrow
\theta'_1$, $\theta_2=0 \rightarrow \theta'_2$ and $\theta_4=0
\rightarrow \theta'_4$,
\begin{eqnarray}
& & {\cal A}_4^{NSR}(\theta'_1,\theta'_3,\theta'_4) \equiv
-\frac{2g_{YM}^2}{\alpha'^2} \langle V'(z'_1,\theta'_1)
\oint W V'(z'_3,\theta'_3) V'(z'_4,\theta'_4) \rangle \nonumber \\
& = & [(D_{\theta_1} \theta'_1)^{-2\alpha' k_1^2} (D_{\theta_3}
\theta'_3)^{-2\alpha' k_3^2} (D_{\theta_4} \theta'_4)^{-2\alpha'
k_4^2}]|_{\theta_1=\theta_3=\theta_4=0} {\cal A}_4^{NSR}(0,0,0) .
\end{eqnarray}
Using the equation of motion (\ref{motion equ}), 
$${\cal
A}_4^{NSR}(\theta'_1,\theta'_2,\theta'_3) = {\cal
A}_4^{NSR}(0,0,0)$$
Then we see the result in (\ref{amp of NSR})
is exactly the 4-point tree amplitude for any values of parameters
$\theta_1$, $\theta_3$ and $\theta_4$.

Since there is no tachyon in the superstring, it is not surprising
that the amplitude doesn't give the terms associated with tachyon poles in
(\ref{boson amp}).

Expanding the function $\frac{\Gamma(-\alpha' s) \Gamma(-\alpha'
t)}{\Gamma(1- \alpha' s - \alpha' t)}$ as in (\ref{gamma
expansion}), the leading terms correspond to the quadratic
Yang-Mills action (\ref{YM interaction}) as in the bosonic
case. The absence of $O(\alpha')$ agrees with the absence of $F^3$
terms in the super Yang-Mills action. The $O(\alpha'^2)$ terms
represent the higher-derivative $F^4$ action in (\ref{power4
interaction}).
The complete action for the effective Yang-Mills theory is then given by eq.\ (\ref{superaction}).

\section*{Acknowledgement}

This work is supported in part by National Science Foundation
Grant No.\ PHY-0354776.

\end{document}